# A SIMPLIFIED PICTURE FOR Π ELECTRONS IN CONJUGATED POLYMERS : FROM PPP HAMILTONIAN TO AN EFFECTIVE MOLECULAR CRYSTAL APPROACH

S. Pleutin* and J-L Fave

Groupe de Physique des Solides (UMR 75-88), Université Paris 7 et 6,

Tour 23, 2 Place Jussieu, 75251 Paris cedex 05, France

* Present address : Max-Planck-Institut für Physik Komplexer Systeme,

Nöthnitzer Strasse, 38 D-01187 Dresden

e-mail : pleutin@mpipks-dresden.mpg.de, fave@gps.jussieu.fr

**Abstract**

An excitonic method proper to study conjugated oligomers and polymers is described and its applicability tested on the ground and first excited states of trans-polyacetylene, taken as a model. From the Pariser-Parr-Pople Hamiltonian, we derive an effective Hamiltonian based on a local description of the polymer in term of monomers; the relevant electronic configurations are build on a small number of pertinent local excitations. The intuitive and simple microscopic physical picture given by our model supplement recent results, such as the Rice and Garstein ones. Depending of the parameters, the linear absorption appears dominated by an intense excitonic peak.

Keywords:

trans-polyacetylene, conjugated polymer, Pariser-Parr-Pople Hamiltonian, exciton states, effective Hamiltonian, linear absorption, electronic correlation, molecular exciton

## I– Introduction

π electrons confer to conjugated polymers attractive electronic properties which are associated to the traditional elastic ones of saturated polymers. For instance, the optical absorption thresholds of these compounds are in the visible or near UV range. Some of them possess a strong and fast non–linear response. Last, some compounds present an insulator–metal transition under doping. During the past twenty years it became clear that the electron–phonon and the electron–electron interactions are both important to correctly describe the lowest electronic excitations of π electrons [1]. Moreover the long-range part of the electron–electron interaction term has somehow to be considered, in order to properly describe, for instance, the excitonic states clearly observed in polydiacetylenes (PDA). The simplest and more studied model Hamiltonian which includes these specific interaction terms is the so-called Pariser–Parr–Pople Hamiltonian [2]

$$H_{ppp} = \sum_{n,\sigma} t_{n,n+1} ( a^{+}_{n\sigma} a_{n+1,\sigma} + a^{+}_{n+1\sigma} a_{n\sigma} ) + \frac{1}{2} \sum_{n,m,\sigma,\sigma'} V_{n,m} ( a^{+}_{n\sigma} a_{n\sigma} - \frac{1}{2} )( a^{+}_{m\sigma'} a_{m\sigma'} - \frac{1}{2} ) \quad (1)$$

In this expression $a^{+}_{n\sigma}$ ($a_{n\sigma}$) are the creation (annihilation) operators of an electron in site $n$ with spin $\sigma$. The first term describes the kinetic energy of the π electrons and their interactions with the atomic cores, expressed using first neighbours hopping integrals. The second term is the Coulomb repulsion between π electrons localized on the sites n and m. Various parametrizations of the coulomb term exist in the literature [1]. Here in view to compare with the results of Yu *et al.*, we adopt their parametrization [3]

$$V_{nm} = U \text{ if } n = m \text{ and } V_{nm} = \frac{V}{|n-m|} \text{ if } n \neq m. \qquad (2)$$

The electron–phonon interaction is treated semi–classically by introducing a linear dependence over the bond lengths in the hopping integrals. In the simplest case, the polyacetylene (PA), an usually adopted parametrization was introduced by Su, Schrieffer and Heeger (SSH) [4, 5]

$$t_{n,n+1} = t_0 + \alpha(u_n - u_{n+1}) \qquad (3)$$

where $\alpha$ is the electron–phonon coupling constant and $u_n$ is the displacement co-ordinate of the nucleus $n$ along the molecular axis. The bond alternation observed in PA is easily reproduced by setting $u_n = (-1)^n u$, yielding a pair of integrals $\beta_d$ and $\beta_s$, the hopping integrals associated with the double (1.35Å) and the single bonds (1.45Å) respectively.

$H_{PPP}$ is invariant under electron–hole transformation. Consequently, the eigenstates of this Hamiltonian can be classified following the electron-hole symmetry classes i.e. the so called (+) and (-) classes. This symmetry allows to simplify the calculations.

*Overview of the methods employed for the study of $H_{PPP}$*. With the parameter values relevant for conjugated polymers, the coulomb terms are approximately of the same order of magnitude than the kinetic ones. In this case, the study of the PPP Hamiltonian becomes a very difficult problem. Only the shortest polyenes (with the number of double bonds N, *$N \leq 8$*) can be properly studied by the usual quantum chemistry methods [2]. For the thermodynamic limit additional, and often drastic, approximations must be applied. The mean field theories [4, 5] gave a simple physical picture of conjugated polymers but, since electronic correlations are discarded, the results are always quantitatively incorrect and even, in some cases, qualitatively wrong [1]. A direct refinement of these methods are the excitonic ones, introduced for instance by Abe *et al.* [6]; using the valence and conduction bands obtained by mean field theory, they perform a configuration interaction including every monoexcited Slater's determinant. For a given parameters range, exciton states may split out of the electron–hole pairs continuum. With such a method the optical absorption spectrum of PDA, in which exciton states have been experimentally observed, is rather well reproduced [6]. However, electronic correlations are still missing; a resulting failure can be found in the loss of size consistency when biexcited states are introduced [7]. Recently Yu *et al.* [3] combined the former theory [6] with a band calculation using the projection technique of Becker and Fulde [8, 9] to include correlation effects. They study the first excitations of the polyparaphenylenevinylene (PPV) considered as an effective linear chain. Good agreement is found with experimentally known energies of the first singlet and triplet excitons as well as with the threshold of the conduction band. Moreover, inclusion of biexcited states brings no size consistency problems contrary to the previous methods. However, with the procedure of renormalization of Yu *et al.*, the two particles of hole–electron pair are independently renormalized. This procedure becomes questionable when the two quasi particles are constrained to stay close to each other as in the exciton states of conjugated polymers. We will examine more precisely this point later.

*Molecular crystal approach*. On the other hand, a more intuitive way is to consider the polymer chain as an one-dimensional molecular crystal of monomers. This approach was first proposed by Simpson [10], refined by many authors [11-13] and recently used by Rice and Garstein [14-16] on polyparaphenylene (PPP) and PPV. Grounds can be found in the prevalent simple bond character of the intermonomer bonding. π electrons are then preferentially confined on the monomers according to their specific topology [17]. In these simple models photoexcitations of the polymers are derived from the local excitations of the monomers and from charge transfer excitations between monomers. Analytical results can then be obtained and a good agreement is found with the absorption spectrum of PPP, after adjustment of the parameter values [16]. The same authors considered also the PPV as a PPP with breaking of the electron–hole symmetry; again, a good agreement is found for the linear absorption [18]. However, the Rice and Gartstein's model (RG) is less microscopic than the one described by $H_{PPP}$. Moreover the field of application of the RG's model is limited to the study of optical absorption spectrum although Mukhopadhyay and co–workers have recently employed a similar molecular exciton method in view to describe the spin–charge cross-over in dimerized chain [19].

*Scope and plan*. In this paper, we present a new method of calculation for the ground state and the first excited states, lending a particular attention to the linear absorption. This new method aims to bridge the gap between the traditional quantum chemistry or solid state physics methods based on $H_{PPP}$ [2, 3] and the simpler molecular exciton methods [14]. Purposely, the electronic configurations of the polymer are built on the basis of localized self-consistent orbitals of the monomers. This specific choice of basis is a natural one for molecular exciton methods. Then from $H_{PPP}$ we build an effective Hamiltonian by selecting the particular subspaces of electronic configurations relevant to the ground state and the primary excitations. With this procedure, the excited states are eventually obtained with the same formal expression as in the RG model. However, with our method, the empirical parameters of RG model are expressed in function of the ones of $H_{PPP}$ and the physical understanding of the states is ameliorated as we will see below. Also, our calculations save size–consistency as the ones of ref [3] but remain much more simple. For simplicity we develop here this new method for the simplest conjugated polymer, the trans–PA in the neutral state; extensions to more complex compounds as PPP and PPV are straightforward and will be presented elsewhere.

The outline of this paper is the following. In section II we introduce the self-consistent orbitals of the monomer (ethylene) and the interaction terms corresponding to this special choice of one-electron functions. We present then the configuration subspace which will be used as model space for the ground state (section III). An approximate diagonalization into this subspace will be given in section IV. Finally we describe in section V the first excited states of (+) electron–hole symmetry - relevant for the linear absorption - and we will discuss our results.

## II- Description of the polymer from the self-consistent orbitals of the monomers

In first approximation conjugated polymers are quasi–one dimensional compounds with a carbon backbone characterised by several types of bonds. For instance, an alternation of the lengths of the different bonds, double and single, occurs in PA, the simplest conjugated polymer. The double bond (1.35Å) is shorter than the single one (1.45Å) and, due to the linear dependence of the hopping terms, $|\beta_d| > |\beta_s|$. Because of this general feature, and in the spirit of the molecular exciton methods, we choose the self-consistent orbitals of the monomers as one-electron base functions. This choice of representation for the polymer is the first step towards an approximate description of the ground state and of the first excited states. The critical parameter which controls the relevance of this particular representation is the bond alternation parameter $z = \frac{a}{\xi}$ where a is the projection of the average C–C spacing onto the chain axis and $\xi$ is the electronic coherence length defined as $\xi = \frac{t_0}{2\alpha u}a$. When $z = 0$ ($\xi \to \infty$), the chain would not be dimerized, with equal lengths for every bond. This case is the more inappropriate to be dealt with our model. Beneficially, larger the order alternation parameter is, and more our particular basis will constitute a good starting point for the study of the π electronic properties. When $z = 1$ ($\xi = a$) the chain is totally dimerized, no electron transfer occurs from a monomer to another and our special choice is the best one.

For the ethylene simple symmetry considerations give the expressions of the self consistent orbitals. We associate with these two orbitals the following creation (destruction) operators for the ethylene $n$

- $B_{n\sigma}^+ = \frac{1}{\sqrt{2}}(a_{2n\sigma}^+ + a_{(2n+1)\sigma}^+)$  $(B_{n\sigma} = \frac{1}{\sqrt{2}}(a_{2n\sigma} + a_{(2n+1)\sigma}))$ associated with the bonding orbital ;

- $A_{n\sigma}^+ = \frac{1}{\sqrt{2}}(a_{2n\sigma}^+ - a_{(2n+1)\sigma}^+)$  $(A_{n\sigma} = \frac{1}{\sqrt{2}}(a_{2n\sigma} - a_{(2n+1)\sigma}))$ associated with the antibonding orbital.

A polyene with N double bonds is then represented by N two levels systems (Fig. 1).

On this basis, the energy terms of $H_{PPP}$, can be reorganised in three different classes (Fig. 2):

(i) The kinetic term which allows an electron or a hole to hop from a monomer to one of its first neighbours. This term is given by $\beta_s$ multiplied by a constant which depends on the topology of the system (and of the orbital symmetries). This constant is equal to 1/2 for the polyacetylene.

(ii) The intramonomer coulomb term $\frac{U-V}{2}$ which introduces electronic correlation inside the monomer, by coupling fundamental configuration of a monomer with the doubly excited one.

(iii) The intermonomer coulomb dipolar terms are of two types:

– First the transition dipole–transition dipole interaction terms between two monomers distant by r monomer units. One effect of these terms is to allow the interaction of intramonomer monoexcitations. The corresponding expression is

$$\Gamma(r) = -\frac{1}{2}\left(V_{1,2r} - 2V_{1,2r+1} + V_{1,2r+2}\right) \qquad (4)$$

– Second the transition dipole–permanent dipole interaction which is expressed by

$$T(r) = -\frac{1}{2}\left(V_{1,2r+2} + V_{1,2r-2}\right) \qquad (5)$$

These dipolar interaction terms decrease very rapidly on r and can be considered as important only for r=1.

In their model Hamiltonian, Rice and Garstein have introduced some semi–empirical parameters which are not considered in the traditional mean field theory [14-16] :

– the "correlation energy gap" $U_{RG}$ which is the energy required in the limit $\beta_s \to 0$ to dissociate a local monomer excitation into a well separated electron and hole;

– the dipole–dipole interaction $V_{RG}$ between local excitations on neighbouring monomers.

With our formulation of the Pariser–Parr–Pople Hamiltonian, for the case of the polyacetylene these semi–empirical parameters acquire a more microscopic meaning. The identification of $V_{RG}$ with $\Gamma(1)$ is immediate but the interpretation of the $U_{RG}$ term is less direct and will be given in section V.

**III– Generative local electronic configurations – Collective excitations**

*Generative local configurations.* In our localized picture of the polyacetylene, electronic configurations will appear as combinations of local electronic configurations. To generate a tractable model, a subset of the Hilbert space has to be selected. A simple way to do it is to retain only some relevant local electronic configurations. In this chapter, we will illustrate this selection procedure in building the ground state subspace. For excited states it will be done in section V.

The ground state traditionally adopted by the molecular excitonic method is [12, 19]

$$|0\rangle = \prod_{n=1}^{N} B_{n\uparrow}^{+} B_{n\downarrow}^{+} |Vacuum\rangle \qquad (6)$$

where $|Vacuum\rangle$ is the vacuum state, i.e. the state without any $\pi$ electron. The energy of this state will serve as reference in the following. The state $|0\rangle$ only includes one type of Local Configuration (LC). We will name it $F$-LC and the corresponding creation operator is

$$F_n^{+} = B_{n\uparrow}^{+} B_{n\downarrow}^{+} \qquad (7)$$

At this level of approximation, the monomers are considered as independent and each monomer possesses 2 $\pi$ electrons which are described in the mean field approximation. With this very simple picture the dynamics of the $\pi$ electrons would be poorly described. In particular, the conjugation phenomenon proper to the $\pi$ systems is not reproduced and the electronic correlations are not introduced inside each monomer. We then improve this description by introducing two electronic local configurations able to interact directly with the state $|0\rangle$ by one of the characteristic interaction terms described above.

1– The $D$–LC in which the monomer $n$ is doubly excited ; this LC is associated with the creation operator :

$$D_n^+ = A_{n\uparrow}^+ A_{n\downarrow}^+ B_{n\uparrow} B_{n\downarrow} \qquad (8)$$

This LC interacts with the state $|0\rangle$ by $\dfrac{U-V}{2}$ and introduces intramonomer electronic correlation; the corresponding energy is $E_D = 4\beta_d$.

2– The $Tc_1^-$–LC in which one electron is transferred from monomer $n$ (or $n+1$) to monomer $n+1$ (or $n$); the transfers on the left and on the right are combined with a (-) sign. This particular linear combination belongs to the class of electron–hole symmetry noted (-) which is the ground state symmetry class. This LC is associated with the creation operator:

$$Tc_{\underline{n}}^+ = \frac{1}{2}\left(A_{n+1\uparrow}^+ B_{n\uparrow} + A_{n+1\downarrow}^+ B_{n\downarrow} - A_{n\uparrow}^+ B_{n+1\uparrow} - A_{n\downarrow}^+ B_{n+1\downarrow}\right). \qquad (9)$$

This LC directly interacts with the state $|0\rangle$ through $\beta_s$ and introduces short range intermonomer charge fluctuations. This effect is intended to correct the too strong localization of the electrons on the double bonds associated with $|0\rangle$; the energy of this LC is $E(1) = 2\beta_d + V + A(1)$ where $A(r)$ is the attractive interaction between the hole and the electron at a distance r

$$A(r) = -\frac{1}{4}\left(V_{1,2r} + 2V_{1,2r+1} + V_{1,2r+2}\right) \text{ (r>0)} \qquad (10)$$

These two kinds of local excited configurations bring about a local improvement of the π electrons dynamics. Therefore in order to improve the dynamics of the whole π electronic system, it is necessary to consider electronic configurations containing a number of $Tc_1^-$–LCs and $D$–LCs, able to introduce simultaneously intermonomer electron delocalization and intramonomer coulomb correlation in any location of the chain. Indeed, we have to build electronic configurations by combining the three LCs $F$–LC, $D$-LC and $Tc_1^-$–LC, which, in this sense, will be called generative local configurations (GLC).

A given electronic configuration involves $n_t$ $Tc_1^-$–LC located on sites labelled $\{x(i)\}$ ($i = 1,..,n_t$) and $n_D$ $D$–LC located on sites labelled $\{y(j)\}$ ($j = 1,..,n_D$); the remaining monomers are represented by $F$–LC

$$|x(1),...,x(n_t); y(1),...,y(n_D)\rangle = Tc^+_{\underline{x}(1)}...Tc^+_{\underline{x}(n_t)}D^+_{y(1)}...D^+_{y(n_D)}|0\rangle. \quad (11)$$

More precisely, a $Tc_1^-$–LC configuration extends on two next neighbour sites; the label $x(i)$ is defined for the left site. In (11) spatial overlap between GLCs are forbidden, so that $x(i) \neq x(j), x(j) \pm 1$ $\forall (i,j)$ and $y(j) \neq x(i), x(i)+1$ $\forall (i,j)$.

The energy of these electronic configurations is given by the following expression:

$$E(\{x(i); y(j)\}) = n_t E(1) + n_D E_D + W(\{x(i)\}). \tag{12}$$

Because of the long range part of the coulomb potential, the energy of an electronic configuration depends on the relative positions of the $n_t$ $Tc_1^-$–LC through the term $W(\{x(i)\})$. This dependence is in fact very smooth so that in practice we will neglect it. We can write :

$$E(\{x(i); y(j)\}) \approx E(n_t, n_D) = n_t E(1) + n_D E_D \tag{13}$$

*Collective excitations.* From the electronic configurations (11), we build collective excitations which are characterised by $\{n_t, n_D\}$, respectively the numbers of $Tc_1^-$–LCs and $D$–LCs. $n_t$ and $n_D$ are not really independent each other; indeed, once the $n_t$ $Tc_1^-$–LCs have been localized on the chain ($n_t = 0,\ldots,N/2$), it is obviously not possible to place a $D$–LC on the sites labelled $\{x(i)\}$ or $\{x(i)+1\}$. Consequently $n_D = 0,\ldots,(N-2n_t)$. For the thermodynamic limit the relevant collective excitations are :

$$|\{n_D, n_t\}\rangle = \frac{1}{\sqrt{N(n_D, n_t)}} \sum_{\substack{\{x(1),\ldots,x(n_t)\} \\ \{y(1),\ldots,y(n_D)\}}} |x(1),\ldots,x(n_t); y(1),\ldots,y(n_D)\rangle. \tag{14}$$

Where $N(n_D, n_t)$ is the number of electronic configurations with $n_t$ $Tc_1^-$–LCs located on sites $\{y_t\}$ and $n_D$ $D$–LCs on sites $\{x_D\}$. In these expressions the summation is over the whole electronic configurations that it is possible to perform. This summation is symbolised by { }.

In the expression (14) the determination of the normalisation constant is purely a problem of enumeration. First let us deal $n_t$ $Tc_1^-$–LCs on the N double bonds of the polymer. Every $Tc_1^-$–LC occupy two neighbour monomer sites, on the both sides of a single bond. The problem is equivalent to place $n_t$ $Tc_1^-$–LCs on (N-1) single bonds but with the additional constraint to avoid their overlap. So, it is necessary to introduce between two $Tc_1^-$–LCs a forbidden zone which extents at least over one single bond. With this topological constraint $n_t$ independent LCs have to be placed on an effective chain of $\tilde{N}$ effective sites. $\tilde{N}$ is equal to the number of single bonds (N-1) minus the number of forbidden bonds $(n_t - 1)$; so $\tilde{N} = N - n_t$ and the configuration number for $n_t$ $Tc_1^-$–LCs is :

$$\mathcal{N}(n_t Tc1-LC) = C_{n_t}^{\tilde{N}} = C_{n_t}^{N-n_t}. \tag{15}$$

Note that this enumeration can be also obtained by recursion.

It is simpler to introduce the on–site $D$–LCs. The problem is to place independently $n_D$ $D$–LCs $(n_D = 0,...,N-2n_t)$ on the remaining (N-2$n_t$) monomers. This is a classical result

$$\mathcal{N}(n_D D;(N-2n_t)) = C_{n_D}^{N-2n_t}. \tag{16}$$

Following (15) and (16) we finally complete the description of the collective excitation

$$N(n_D, n_t) = C_{n_t}^{N-n_t} C_{n_D}^{N-2n_t}. \tag{17}$$

**IV– Approximate ground state**

Generally speaking, determining the exact ground state of $H_{PPP}$ is a very difficult task which can only be performed numerically for finite compounds with less than six double bonds. In the polymer limit, various more or less drastic approximations have been proposed such as the simplified ground state of the molecular exciton methods (6) [12, 13, 19], the traditional mean field ground state [6] or the sophisticated Gützwiller variational solution of the Peierls–Hubbard Hamiltonian [20]. In this section an approximate ground state is built by diagonalizing $H_{PPP}$ in the Hilbert's subspace spanned by the collective excitations $|\{n_D, n_t\}\rangle$. We believe that this approximation grasps the essential features of the ground state in view to describe the linear spectroscopic properties of conjugated polymers.

The collective excitations expressed by (14) and (17) interact each other through two distinct interaction terms :

$$I_t(n_t; n_D) = \langle \{n_D, n_t + 1\} | H_{PPP} | \{n_D, n_t\} \rangle \qquad (18\text{-a})$$

$$I_D(n_D; n_t) = \langle \{n_D + 1, n_t\} | H_{PPP} | \{n_D, n_t\} \rangle \qquad (18\text{-b})$$

the intermonomer delocalization interaction term and the intramonomer correlation interaction term respectively. The dipolar terms which have been described above are small enough to be neglected in this approximate treatment of the ground state.

*Intramonomer electronic correlation.* First we consider the configuration subspaces which are spanned by the collective excitations $|\{n_D, n_t\}\rangle$, such as $n_D$ varies from $0$ to $(N - 2n_t)$; each subspace is characterised by a specific value of $n_t$. Only the second interaction term acts inside a given subspace :

$$I_D(n_D; n_t) = \sqrt{(n_D + 1)(N - 2n_t - n_D)} \frac{U - V}{2} \qquad (19)$$

where $n_D = 0, \ldots, (N - 2n_t)$, that allows an easy diagonalization of $H_{PPP}$ in these subspaces. Indeed, from a particular electronic configuration with $n_t$ localized $Tc_1^-$–LCs $|x(1), \ldots, x(n_t)\rangle$, we can independently introduce intramonomer electronic correlation on each of the $(N - 2n_t)$ remaining double bonds [21]. We then obtain for each monomer two states $|-\rangle$ and $|+\rangle$ associated with the energies $\varepsilon_-$ and $\varepsilon_+$ respectively

$$\varepsilon_{\pm} = 2\beta_d \pm \frac{1}{2}\sqrt{16\beta_d^2 + (U-V)^2}. \qquad (20)$$

If the state $|0_n\rangle$ in which the particular monomer n is in its own ground state is introduced, the states $|-\rangle$ and $|+\rangle$ are written as

$$\begin{cases} |-\rangle = a|0_n\rangle + bD_n^+|0_n\rangle \\ |+\rangle = aD_n^+|0_n\rangle - b|0_n\rangle \end{cases} \text{ where } a = \frac{(U-V)}{\sqrt{4\varepsilon_-^2 + (U-V)^2}} \text{ and } b = \sqrt{1-a^2}. \qquad (21)$$

From the electronic configuration $|x(1),\ldots,x(n_t)\rangle$, we obtain a wavefunction incorporating intramonomer electronic correlation, $|x(1),\ldots,x(n_t)\rangle^c$ by associating to each monomer not implied in a $Tc_1^-$–LC, a state such as $|-\rangle$:

$$|x(1),\ldots,x(n_t)\rangle^c = \sum_{j=0}^{N-2n_t} \sum_{\substack{\{y(1),\ldots,y(j)\} \\ y(i)\neq x(1),\ldots,x(n_t)}} a^{N-2n_t-j} b^j D_{y(1)}^+ \ldots D_{y(j)}^+ |x(1),\ldots,x(n_t)\rangle. \qquad (22)$$

It is straightforward to verify that these states are eigenvectors of $H_{PPP}$ in the subspace spanned by the collective excitations $|\{n_D,n_t\}\rangle$ with a fixed number $n_t$.

*Intermonomer electronic delocalization.* From these $|x(1),...,x(n_t)\rangle^c$ we build new collective excitations including intramonomer electronic correlation, noted $|n_t\rangle^c$, for $n_t$ $Tc_1^-$–LCs :

$$|n_t\rangle^c = \left(C_{n_t}^{N-n_t}\right)^{-1/2} \sum_{\{x(1),...,x(n_t)\}} |x(1),...,x(n_t)\rangle^c = \sum_{j=0}^{N-2n_t} a^{N-2n_t-j} b^j \sqrt{C_j^{N-2n_t}} |\{j,n_t\}\rangle \quad (23)$$

The energies associated to these collective excitations from $|0\rangle^c$ are given by

$$E^c(n_t) = n_t(E(1) - 2\varepsilon_-) = n_t(E(1) - 2\varepsilon_c). \quad (24)$$

These collective excitations interact through the following terms

$$I_t^c(n_t) = {}^c\langle n_t+1|H_{PPP}|n_t\rangle^c = \left[(n_t+1)\frac{(N-2n_t)(N-2n_t-1)}{N-n_t}\right]^{1/2} a^2 \beta_s. \quad (25)$$

Because of the topological constraint typical of the $Tc_1^-$–LCs, the mathematical problem which is governed by the interaction term (25) and the energy term (24) is not analytically solvable, contrary to the problem of the intramonomer electronic correlation. To simplify further we perform an additional approximation in which the interaction term $I_t^c$ is replaced by a simplest one analogous to (19)

$$J_t^c(n_t) = \left[(n_t+1)((N-1)/3 - n_t)\right]^{1/2} \sqrt{3}a^2\beta_s. \quad (26)$$

This interaction term $J_t^c$ constitutes an excellent approximation $I_t^c$ for the low values of $n_t$ (Fig. 3). In the following we will check that the collective excitations $|n_t\rangle^c$ with $n_t$ greater than approximately N/4 are not relevant for the description of the ground state, so that the approximation holds even in an unfavourable case.

With the interaction term $J_t^c$ the problem of intermonomer delocalization becomes formally similar to the one of intramonomer electronic correlations. For the later, we have seen that the system behaves as N coupled independent two levels systems, the coupling being (U-V)/2. Here, the problem is reduced to (N-1)/3 effective independent two levels systems $|a\rangle$ and $|b\rangle$. These states have the energies $E_a = 0$ and $E_b = E(1) - 2\varepsilon_c$ and are coupled by the effective interaction $\sqrt{3}a^2\beta_s$. For the polymer ground state we have only to consider the lowest eigenstate in energy, noted $|-\rangle^t$. This state can be easily expressed :

$$|-\rangle^t = a_t|a\rangle + b_t|b\rangle \quad (27)$$

which is associated with the energy

$$\varepsilon_t = \frac{E_1 - 2\varepsilon_c}{2} - \frac{1}{2}\sqrt{(E_1 - 2\varepsilon_c)^2 + 12\, a^4 \beta_s^2} \qquad (28)$$

where

$$a_t = \frac{\sqrt{3}\, a^2 \beta_s}{\sqrt{\varepsilon_t^2 + 12\, a^4 \beta_s^2}} \quad \text{and} \quad b_t = \sqrt{1 - a_t^2}\,. \qquad (29)$$

For convenience let's define $\tilde{N}_t = \boldsymbol{E}((N-1)/3)$ where $\boldsymbol{E}$ takes the integer part. With the help of (27) we finally obtain an approximate description of the ground state :

$$|GS\rangle = \sum_{n_t=0}^{\tilde{N}_t} a_t^{\tilde{N}_t - n_t} b_t^{n_t} \sqrt{C_{n_t}^{\tilde{N}_t}}\, |n_t\rangle^c \qquad (30)$$

associated with the energy

$$E_{GS} = N\varepsilon_c + \frac{N-1}{3}\varepsilon_t. \qquad (31)$$

In order to measure the quality of our analytical approximation, we have performed a numerical calculation with the "exact" interaction term (25). In the figure 4, we have reproduced the square of the coefficients of the ground state wave function on the basis of the collective excitations $|n_t\rangle^c$ calculated for $z = 0$ and in neglecting the coulomb terms for a long polyene. This parameter value corresponds to the most unfavourable choice to test our method. We can see that for the ground state wave function the gaussian approximation is indeed very good. Moreover, only the collective excitations $|n_t\rangle^c$ with a low $n_t$ appear to be important. The corresponding energy too agrees very well (to 1%).

*Comparison with the exact result of the SSH Hamiltonian.* The Pariser–Parr–Pople Hamiltonian admits analytical solutions if we neglect the electron–electron interaction term, to obtain the so called Su–Schrieffer–Heeger Hamiltonian [5]. In the following we compare our approximate result with the exact analytical which is obtained in the framework of the SSH Hamiltonian, versus the bond alternation parameter. In figure 5 we have represented the ratio $\Delta(z) = \dfrac{E_{GS}(z)}{E_{SSH}(z)}$ in which $E_{SSH}(z)$ is the exact analytical result of the ground state energy.

(i) For $z = 1$, our result becomes exact. This is not surprising since in this limiting case the monomers are totally independent.

(ii) For $z = 0$, we obtain almost 92% of the exact result; this good agreement constitutes a surprise since our model seems rather unsuitable to this limit. Indeed this interesting result shows that the most important charge fluctuations are on the range of the intramonomer and of the nearest neighbours, even in the unrealistic case of a conjugated polymer in which all bonds would have the same length.

It is possible to refine further this crude description by considering additional GLCs; the figure 5 shows also the improvement when all the GLCs implying two neighbour sites are taken into account.

For the Polyacetylene, in which the generally admitted parameter values give $z \approx 0,15$ [1, 5], the agreement reaches about 97%. On an other hand, it is possible to describe other compounds as effective linear chains with a more pronounced bond length alternation [3]. For example the polyparaphenylene can be crudely associated to $z \approx 0,3$. In such a case the agreement is excellent, we obtain about 99%. By looking at these good results, we guess that our approximation, so crude it could seem with only three GLCs (*F*–LC, *D*–LC and $Tc_1^-$–LC), keeps the essential features of the ground state.

An evident flaw of our approximation at this level is the neglect of long-range intermonomer charge fluctuation. "Long" means here over the next neighbour monomer. In the SSH model, all the charge transfer LCs (at any distance) are degenerate. Note that, on the contrary, when electron–hole coulomb interaction is taken into account, the energies of the short range LCs are decreased and our approximation will become better.

*Finite size effect*. The approximate intermonomer delocalization energy (IDE) for the polymer fails for the smallest polyenes. For instance for $N = 2$ we obtain with this expression only 80% of the exact result. However, this expression can be easily improved. Indeed, for $N = 2$ the IDE is obviously given by

$$\varepsilon_2 = \frac{E_1 - 2\varepsilon_c}{2} - \frac{1}{2}\sqrt{(E_1 - 2\varepsilon_c)^2 + 4a^4\beta_s^2} \qquad (32)$$

and not by $\frac{\varepsilon_t}{3}$. We can then rewrite the IDE for a polyene with N double bonds including this correction:

$$E_{deloc}(N) = \varepsilon_2 + \frac{(N-2)}{3}\varepsilon_t \qquad (33)$$

With this new expression the IDE of any oligomer is well reproduced with an error always lower than 2%. This improvement of the delocalization energy will become particularly important in the following.

**V– First excited states of (+) electron-hole symmetry - Exciton states**

In this paper we only will consider now the lowest excited states of (+) electron hole symmetry. They are the relevant states in order to study the threshold of the linear absorption. We have generated above a subspace relevant for the ground state and containing only $|\{n_D, n_t\}\rangle$ electronic configurations. For the excited states, we select an other model subspace based on electronic configurations which differ of $|\{n_D, n_t\}\rangle$ only by localized perturbative areas. So, in our formalism, the excited states are composed of two different parts:

(i) A local zone - called the core of the excitation - which is a local perturbation of the ground state system. The description of this local excitation requires new GLCs of (+) electron–hole symmetry. We will examine them below.

(ii) Outside the core of the excitation, the dynamics of the π electrons remains described by $F$, $D$ and $Tc_1^-$–LC as in the ground state.

Let us first introduce the new GLCs on which is based the description of the cores of the excitations. For the states of interest here, these GLCs are monoexcitations of (+) electron-hole symmetry which are expressed by the following creation operator :

$$Tc_n^+(r) = \frac{1}{2}\left(A_{n+r\uparrow}^+ B_{n\uparrow} + A_{n+r\downarrow}^+ B_{n\downarrow} + A_{n\uparrow}^+ B_{n+r\uparrow} + A_{n\downarrow}^+ B_{n+r\downarrow}\right) \tag{34}$$

This operator creates a local mono excitation in which a hole and an electron are r monomers apart. These GLCs will be noted $Tc_r^+$-LC. Their energies depend on $r$ :

$$E(0) = 2\beta_d + \frac{U-V}{2} \tag{35}$$

$$E(r>0) = 2\beta_d + V + A(r). \tag{36}$$

The model subspace for excited states is spanned by the complete set of electronic configurations with one charge transfer excitation of (+) symmetry located on sites $t(n)$ and $t(n+r)$, $n_t$ $Tc_r^-$-LCs located on sites labelled $\{x(i)\}$ ($i=1,..,n_t$) and $n_D$ $D$–LCs located on sites labelled $\{y(j)\}$ ($j=1,...,n_D$). Of course the same constraints as seen for the ground state have to be considered i.e. no spatial overlap of the different LCs can occur. A general expression is then

$$|t(n),t(n+r);x(1),...,x(n_t);y(1),...,y(n_D)\rangle = Tc_n^+(r)Tc_{x(1)}^+...Tc_{x(n_t)}^+ D_{y(1)}^+...D_{y(n_D)}^+|0\rangle$$

(37)

The diagonalization into this model subspace follows the previous approximate treatment performed for the ground state.

The excitations (37) interact via $H_{ppp}$. As for the ground state we begin by introducing intramonomer electronic correlations : we associate a state $|-\rangle$ (21) to every $F$–LC of $|t(n),t(n+r);x(1),...,x(n_t)\rangle$, to obtain

$$|t(n),t(n+r);x(1),...,x(n_t)\rangle^c =$$
$$\sum_{j=0}^{N-2n_t-2} \sum_{\substack{\{y(1),...,y(j)\} \\ y(i)\neq t(n),t(n+r),x(1),...,x(n_t)}} a^{N-2n_t-2-j} b^j D_{y(1)}^+...D_{y(j)}^+ |t(n),t(n+r);x(1),...,x(n_t)\rangle$$

(38)

It is important to notice that, because of the presence of the charge transfer excitation, the number of $F$–LCs is in this case only $N-2n_t-2$ (or $N-2n_t-1$ if the excitation is concentrated onto only one monomer, r=0). Indeed it is not possible to introduce $D$–LCs in the monomer sites occupied by this perturbation. Consequently, this results in a loss of ground state intramonomer correlation energy which then increases the excitation energy [9].

The collective excitations (38) interact via the transfer integral in a similar way to the collective excitations (23) relevant for the ground state. However the problem is here more complex. Indeed the presence of a $Tc_r^+$-LC divides the chain in three different sections:

– the part on the left of the $Tc_r^+$-LC which contains $N_L$ monomers,

– the part on the right of the $Tc_r^+$-LC which contains $N_R$ monomers,

– the part inside the $Tc_r^+$-LC which contains $N_I$ monomers located between $t(n)$ and $t(n+r)$.

We have seen before that the expression (33) is valid (with a very good agreement) for any chain size. The three parts of the chain can be independently solved in the same way than the ground state. So we obtain for the three parts

$$- E_{deloc}^{L/R} = \varepsilon_2 + \frac{(N_{L/R} - 2)}{3} \varepsilon_t \tag{39}$$

$$- E_{deloc}^{I}(N_I) = \varepsilon_2 + \frac{(N_I - 2)}{3} \varepsilon_t, \text{ if } N_I > 1$$
$$= 0 \qquad \text{else.} \tag{40}$$

The total intermonomer delocalization energy is the sum of these three different parts; this is a function of $r$, the distance between the hole and the electron of the excitation $Tc_r^+$–LC.

$$E_{deloc}(r) = (2 + \Theta(r-3))\varepsilon_2 + \frac{N - 6 - \Theta(r-2) - \Theta(r-3)}{3}\varepsilon_t \qquad (41)$$

where $\Theta(x)$ is the Heavyside's function which is equal to zero for x below zero and one for x above zero. In every case there is a loss of IDE in the excited states by respect to the ground state.

The corresponding wave function is given by

$$|t(n);t(n+r)\rangle^{dc} = \sum_{n_t=0}^{\tilde{N}_r} a^{\tilde{N}_r - n_t} b^{n_t} |t(n);t(n+r);\{n_t\}\rangle^c \qquad (42)$$

where $\tilde{N}_r = \mathbf{E}((N_L + N_R + N_I - 3)/3)$ and $|t(n);t(n+r);\{n_t\}\rangle^c$ are collective excitations analogue to (23) but with local excitation on $n$ and $n+r$. In this expression, every of the three different parts introduced by the "defect" $Tc_r^+$-LC is independently described by a state as (30). However this approximation found for the ground state is relevant for the limiting case $N \to \infty$. Consequently, the parts outside the $Tc_r^+$-LC are well described in (42). On the contrary, the description of the part inside the $Tc_r^+$-LC could become inappropriate when the charge transfer extends only over few monomer units; we have neglected these finite size effects. Moreover, in practice we have neglected too the renormalization of the interaction terms due to the effects of intermonomer delocalization. Taking into account these effects is straightforward without significant change in the results.

*Dressing of the excitations.* Yu and co–workers dressed the particles by a polarization cloud following the projection technique [3, 9]. In this work, we dress each $Tc_r^+$-LC using a perturbative treatment. In order to perform this, for a given $Tc_r^+$-LC we consider the local configurations $|I\rangle$ called corrective local configurations (CLC), with excitation energy $E_I$, which interact directly with it through $t_I$. The effect of $|I\rangle$ on the energy of the excitation is then accounted for by a simple second order perturbation expansion which depends on $r$

$$\varepsilon_P(r) = \sum_I \frac{t_I^2}{E(r) - E_I} \qquad (43)$$

The relevant CLCs give two distinct contributions to the polarization cloud:

(i) the first kind of contribution is the dipolar one which is due to the long range part of the Coulomb term;

(ii) the second kind of contribution introduces additional kinetic terms through the transfer integral.

The energy of a local excited configuration, including intramonomer electronic correlations and intermonomer electronic delocalizations, is then given by the following sum

$$\overline{E}^{Dc}(r) = E(r) + (N - 1 - \Theta(r-2))\varepsilon_c + E_{deloc}(r) + \varepsilon_p(r). \tag{44}$$

This energy can be expressed by respect to the ground state energy

$$\varepsilon(r) = \overline{E}^{Dc}(r) - E_{GS} = E(r) - (1 + \Theta(r-1))\varepsilon_c + (1 + \Theta(r-3))\varepsilon_2$$
$$- (3 + \Theta(r-1) + \Theta(r-2))\varepsilon_t / 3 + \varepsilon_p(r)$$

(45)

In this expression three different competing terms contribute to decrease or increase the excitation energy compare to the crude energy $E(r)$. There is a loss of ICE and IDE and a gain of polarization energy.

*Renormalization of the interaction terms.* The effective interaction terms between $Tc_r^+$-LCs are also easily introduced by the quasi–degenerated second order perturbation theory. Indeed, let us note $|1\rangle$ and $|2\rangle$ two $Tc_r^+$-LCs of energies $E'$ and $E''$ respectively. These two $Tc_r^+$-LCs interact with some identical CLC $|I\rangle$ through interaction terms noted $t_{1I}$ and $t_{2I}$. In these conditions an effective interaction exists between $|1\rangle$ and $|2\rangle$, which is expressed by the following simple expression

$$t_{12}^{eff} = \frac{1}{2} \sum_I t_{1I} t_{2I} \left\{ \frac{1}{E' - E_I} + \frac{1}{E'' - E_I} \right\} \tag{46}$$

This correction modifies the crude interaction term between the two LCs $|1\rangle$ and $|2\rangle$.

*Diagonalization.* Once the energy terms (45) and the interaction terms have been determined, we obtain the analogue of the Rice and Gartstein's model for PA but with parameters directly expressible from $H_{PPP}$. We have seen above the simple expression of $V_{RG}$. We can now express the "correlation energy gap" with the help of the $H_{PPP}$ parameters:

$$U_{RG} = \lim_{\beta_s \to 0} (\varepsilon(\infty) - \varepsilon(0)) = V + \varepsilon_p(\infty) - \varepsilon_p(0) = V - \frac{2T(1)^2}{(U-V)} \qquad (47)$$

The excitation spectrum is then easy to calculate. Because the system is translationally invariant we build first, from the local excitations (42), the collective excitations characterised by the wavenumbers $k$:

$$|k,r\rangle = \sqrt{\frac{2}{N-r+1}} \sum_n \sin\frac{k\pi n}{N-r+1} |t(n); t(n+r)\rangle_l^{dc} \qquad (48)$$

The second step consists in diagonalizing the following tridiagonal matrix

$$\langle k,r|H_{PPP}|k',r'\rangle = \delta_{kk'}\{\delta_{rr'}\varepsilon(r)+\delta_{r',r\pm 1}I_{r,r\pm 1}(k)\} \qquad (49)$$

where $\delta$ is the Kronecker's index and $I_{r,r\pm 1}(k)$ the interaction term between $|k,r\rangle$ and $|k,r\pm 1\rangle$.

Because the interactions are local, only the collective excitations with the same wave number $k$ may interact. Here we have to consider only the k=0 subspace in which the whole oscillator strength is concentrated.

*Size-consistency of our procedure*

Excitonic calculations are usually performed at the SCF level via a configuration interaction (CI) in the subspace of monoexcitations [6]. It is well known that size-consistency of such a procedure is not insured and these methods fail for instance when biexcited configurations are considered [7]. This is the case for the study of 2A$_g$ states of some conjugated polymers [1,2]. The reason is indeed that it is impossible to treat the ground state and the excited states on an equal footing: doubly excited configurations introduce some electronic correlation in the ground state; on the contrary, the excited states (1B$_u$, 2A$_g$) stay at an uncorrelated level. Consequently, the excitation energies diverge when the system size increases. This is the case for any incomplete CI procedure not restricted to monoexcited determinants.

In our procedure, the electronic system is described at a local scale. The total energy is an extensive quantity and the only differences between the ground state and the excited configurations (equations 30 and 42) are localized on the core of the excitation. The other part of (42) is treated exactly in the same way as for the ground state. Consequently, our procedure, as the one of the ref [3], is size-consistent. The introduction of higher excitation processes becomes straigthforward: it would suffice to consider some other GLCs for building the corresponding core excitations.

*Comparison with the Yu et al. results*. In view of describing the first excited states of the PPV, Yu *et al.* have considered an effective Pariser–Parr–Pople Hamiltonian [3]. Assuming that the benzene rings just affect the electron transfer between double bonds, they have considered the PPV as a PA with a slightly more pronounced dimerization; they took $z = 0.19$. They performed a calculation of the excitation spectrum in two different steps :

– first, they determined the band structure, taking into account the effect of the electronic correlations by applying the projection technique of Becker and Fulde [8, 9] ;

– second, using the correlated band structure they performed an excitonic calculation following Abe *et al.* [6].

With $z = 0.19$, $U = 3t_0$, $V = t_0$ and $t_0 = 2\,eV$, the lowest singlet exciton of (+) electron–hole symmetry appears at $E_{exc} = 2.42\,eV$ and the edge of the conduction band appears at $E_{gap} = 3.58\,eV$. These calculated values agree well with the experimental ones. With our method the calculation of the excitation spectrum for the same parameters gives $E_{exc} = 3.30\,eV$ and $E_{gap} = 3.55\,eV$. The agreement for the band gap is excellent, contrary to the exciton energy value, and this pinpoints a fundamental difference between the two models. We have to note before going further that the value of $E_{exc}$ is calculated on a finite polyene with N=75 double bonds. However the convergence of the calculation with the system size is fast and the finite size does not affect noticeably the final results. The discrepancy between the two calculations can rather be attributed to the renormalization procedures which are radically different. In the method of Yu *et al.* the hole and the electron are renormalized by the projection technique independently each other [9]. Therefore the hole and electron polarization clouds are invariant by respect to the separation distance r between them. The energies $\varepsilon(r)$ of the charge transfer excitation vary only by the effect of the attractive term between the hole and the electron ($\varepsilon(r) < \varepsilon(r')$ if $r < r'$). This approximation is justified whenever the description in term of bands is sufficient; then electron and hole are far away on the average. In the excitonic states, they are constraint to stay close to each other and, moreover, it is known that the exciton radius is rather small in conjugated polymers. In this extreme case the polarization clouds of the two quasi–particles may interact notably and the approximation of Yu *et al.* becomes questionable. In our method we adopt a completely different procedure. Indeed we have renormalized the local excitation $Tc_r^+$-LC in which a hole and an electron are separated by r monomer units. Then the hole and the electron are renormalized together and the correction $\varepsilon_p(r)$ depends on the distance r between them. When the excitation is a band to band excitation, the two procedures are equivalent because we can consider that the two quasi–particles are (principally) far from each other. So we obtain approximately the same value for $E_{gap}$. On the contrary, for an excitonic state, the two procedures are different and so give different values for $E_{exc}$. Yu *et al.* should have overestimated the gain of polarization energy. Indeed, when the two particles are very close, we can roughly guess that in the intermediate zone the polarization gain is counted twice.

In view to confirm these conclusions, it is easy to adapt our formalism to the hypothesis of Yu *et al.*. In this case we write the excitation energy of a charge transfer excitation of radius r $\varepsilon_y(r)$ as

$$\varepsilon_y(r>0) = \varepsilon(\infty) + A(r) \tag{50}$$

$$\varepsilon_y(0) = \varepsilon(\infty) + \frac{U-3V}{2} + \varepsilon_c \tag{51}$$

where $\varepsilon(\infty)$ is calculated following the expression (45). We have yet to neglect any difference between the interaction terms in the matrix (49) and to take $I_{r,r\pm 1}(k) = I_\infty(k)$, the interaction term between charge transfers of infinite radius. With these new values, we obtain $E_g = 3.55\,eV$ and $E_u = 2.47\,eV$. This excellent agreement completely confirms the origin we have assigned to the discrepancy.

*Exciton versus conduction band absorption*. Let us finally introduce the interaction of an electric field $\vec{E}(\vec{r},t)$ with the $\pi$ electrons via the dipolar approximation

$$H_{int} = e\vec{r}\vec{E}(\vec{r},t) \tag{52}$$

where $e$ is the electron charge. This approximation is justified as long as the wavelength $\lambda$ of the electric field is greater than the characteristic length a; this is effectively the case in the visible range.

The states of different electron–hole symmetry are coupled through $H_{int}$. The linear absorption spectrum at low energy is entirely determined by the monoexcited states that we have presented in this section. In figure 6 we have represented the component along the molecular axis of the calculated oscillator strength for $z = 0.19$, $U = 3t_0$, $V = t_0$ and $t_0 = 2\,eV$. As it is observed experimentally for the PDA, the major part of the oscillator strength is concentrated in the excitonic peak. In counterpart, the continuum of electron–hole becomes quasi–invisible. Results with other parameters show that, larger the binding energy of the exciton is, smaller the range of the prominent charge transfer states and the higher the intensity of the associated transition will be. Similar results have been found very recently but in a strong correlation approximation not really appropriate for conjugated polymers [22].

### VI– Conclusion

Intramonomer transfer integrals are larger than intermonomer ones in conjugated polymers. In this work, we have taken intentionally advantage of this characteristic feature of the conjugated polymers to build the electronic configurations of the polymer from the monomer orbitals. We have then diagonalized the Pariser–Parr–Pople Hamiltonian on a reduced Hilbert space. By doing this, from PPP Hamiltonian, we obtain for the excited states calculations a simple effective molecular exciton method.

With the adopted local description, each polymer electronic configuration is a distinct combination of several local electronic configurations. For the ground state, we have decided to retain only three kinds of local configurations, the so–called generative local configurations : $F$, $D$ and $Tc_1^-$–LC. The first GLC represents a monomer in its ground state; the second and the third ones introduce intramonomer correlation and intermonomer delocalization, respectively. These three GLCs permit to build a configuration subspace, on which we give an approximate analytical solution of $H_{PPP}$. This solution is in good agreement with the exact result of the mean field approximation of the Su–Schrieffer–Heeger Hamiltonian.

Excited states of low energy are then dealt with as local perturbations of the ground state so obtained. The composition of the perturbative zone (the core of the excitation) depends on the nature of the considered excitation. In this paper we only consider the lowest excitation of (+) electron–hole symmetry generated by $Tc_r^+$-LC, in which one electron is transferred from a monomer n to the monomer (n+r). Furthermore, these $Tc_r^+$-LCs are dressed by a perturbative treatment. The resulting excited state energies are determined by the losses of correlation and delocalization by respect to the ground state, as well as by electron-hole attraction and the counteracting polarization energy. The results obtained in this manner possess the characteristics expected for conjugated polymers: the first excitations are excitons of short radius and the oscillator strength is essentially concentrated into these excitonic transitions. Our calculations moreover show the necessity to properly consider the interaction between the two quasi-particles which constitute the exciton, especially when its radius is small.

For simplicity, this approach has been applied here to the neutral trans-polyacetylene. In reason of their topology, other compounds need a more complex description of monomers and a larger number of GLCs. The polymer excited states will stem from the various possible excitations of a monomer. The corresponding collective excitations depend then on their mutual coupling, particularly on the relevant charge transfer integrals, whose some can be vanishing small. Several distinct excitonic states are then possible, besides excitations remaining mainly localized on monomers.

# Figures captions

Figure 1
Representation of a polyene by N two-levels systems

Figure 2
PPP interactions expressed on the basis of monomer self consistent orbitals

Figure 3
Variation of the interaction term between the collective excitation $|n_t\rangle^c$ et $|n_t+1\rangle^c$ versus the number of charge-transfer excitations $n_t$ for a N=200 polyene. $I_t^c$ exact term (25) ; $J_t^c$ approximated term (26).

Figure 4
Weight of the collective excitations $|n_t\rangle^c$ in the ground state of a chain of 168 monomers (without alternation nor electron-electron interaction).
Open circles: numerical result
Solid circles: gaussian approximation

Figure 5
Ratio of the ground state energy (31) to the exact analytical SSH solution versus the bond alternation parameter z
Solid circles: using three GLCs (*F*–LC, *D*–LC and $Tc_1^-$–LC)
Open circles: using in addition all the GLCs implying next-neighbours

Figure 6
Component of the linear absorption along the molecular axis, in arbitrary units.
The exciton peak and the threshold of the conduction band have been calculated for z=0.19, t0=2eV, U=3t0, V=t0

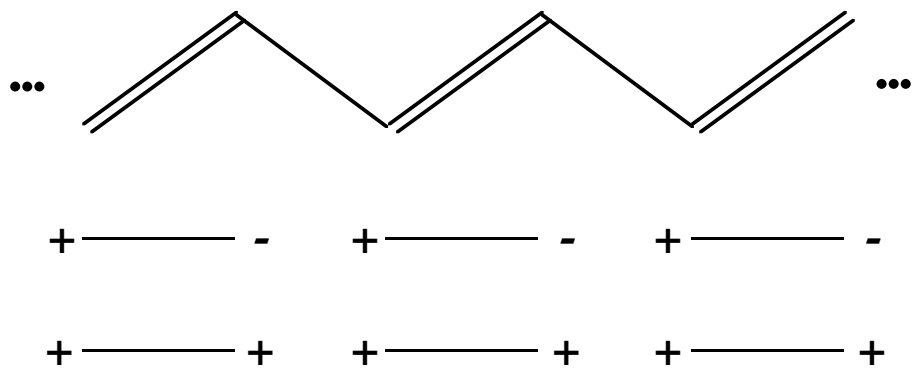

Figure 1

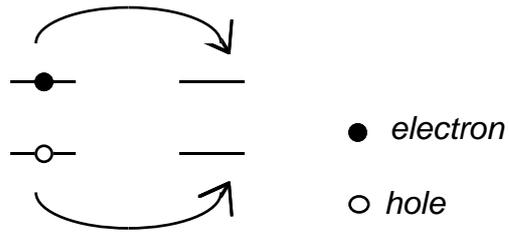

a) Hopping term

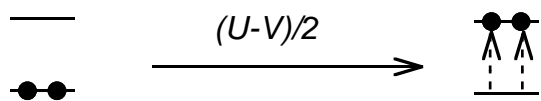

b) Intramonomer coulomb term

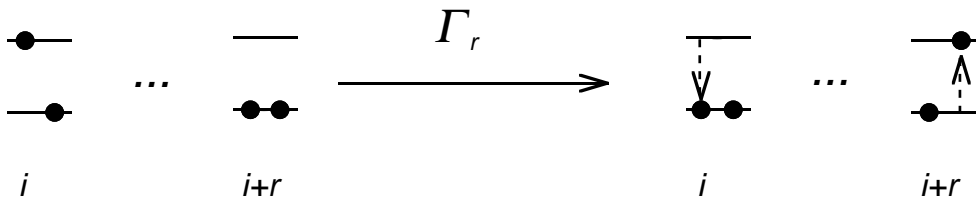

c) Transition dipole-transiton dipole interaction term

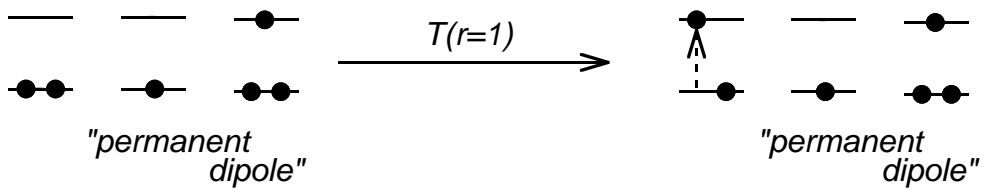

d) Transition dipole–permanent dipole interaction term

Figure 2

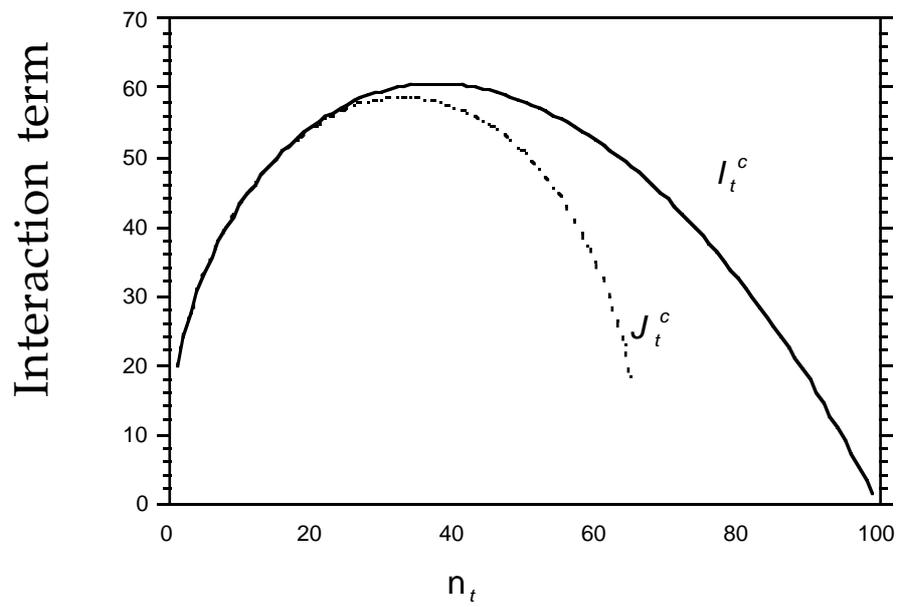

Figure 3

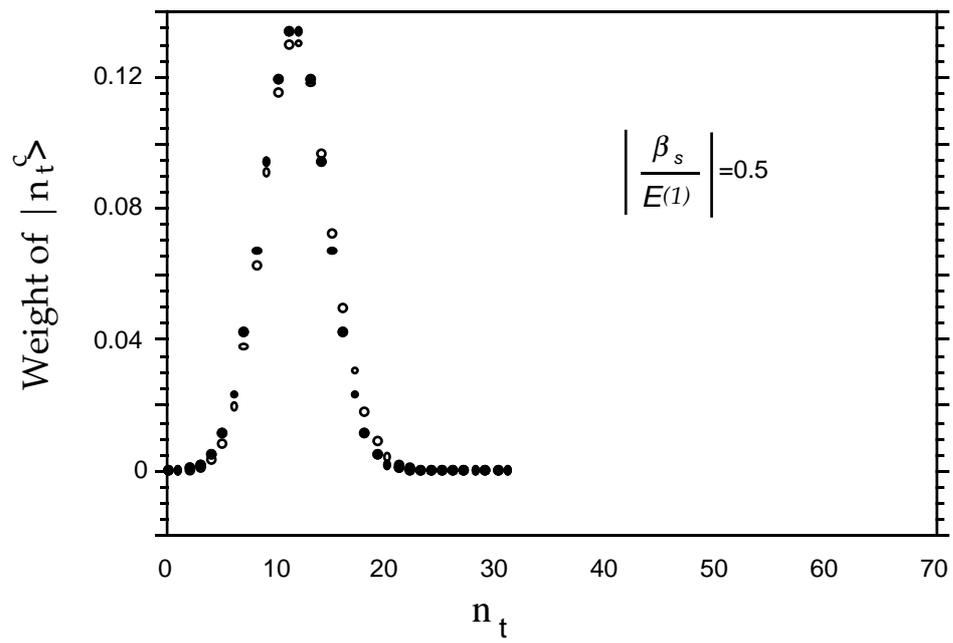

Figure 4

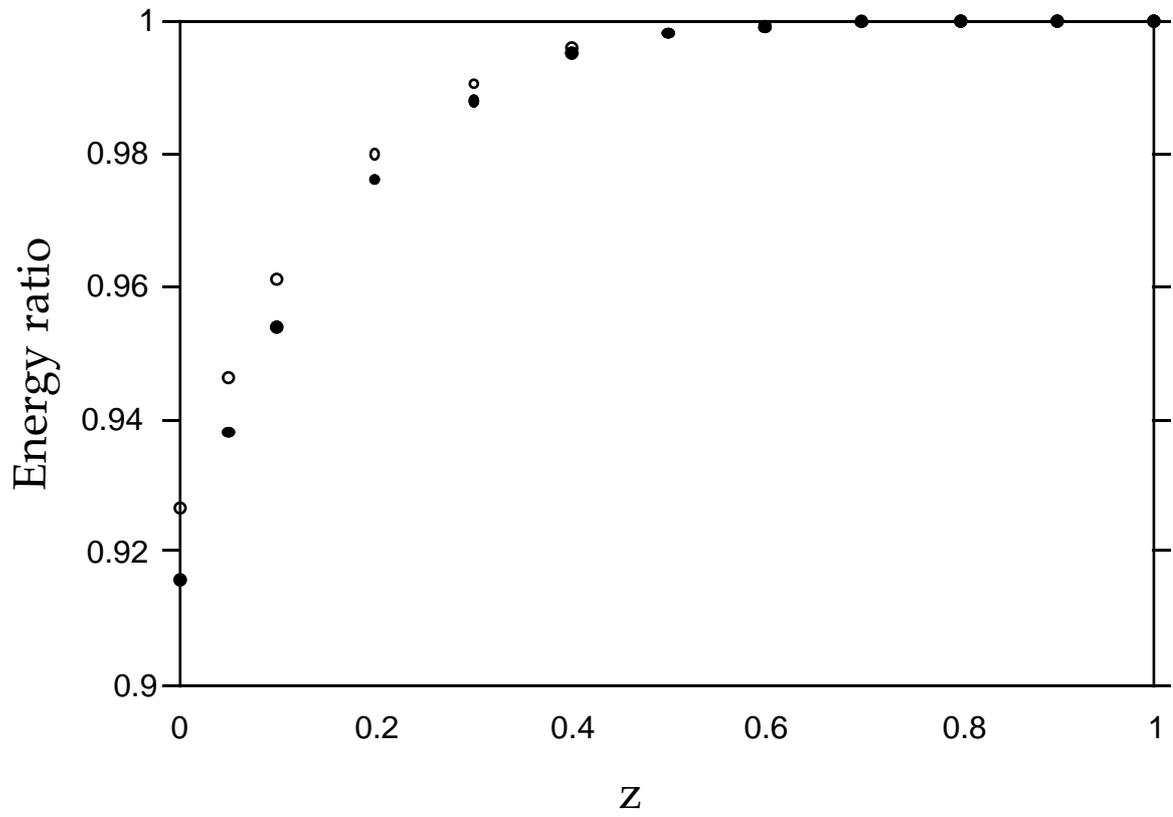

Figure 5

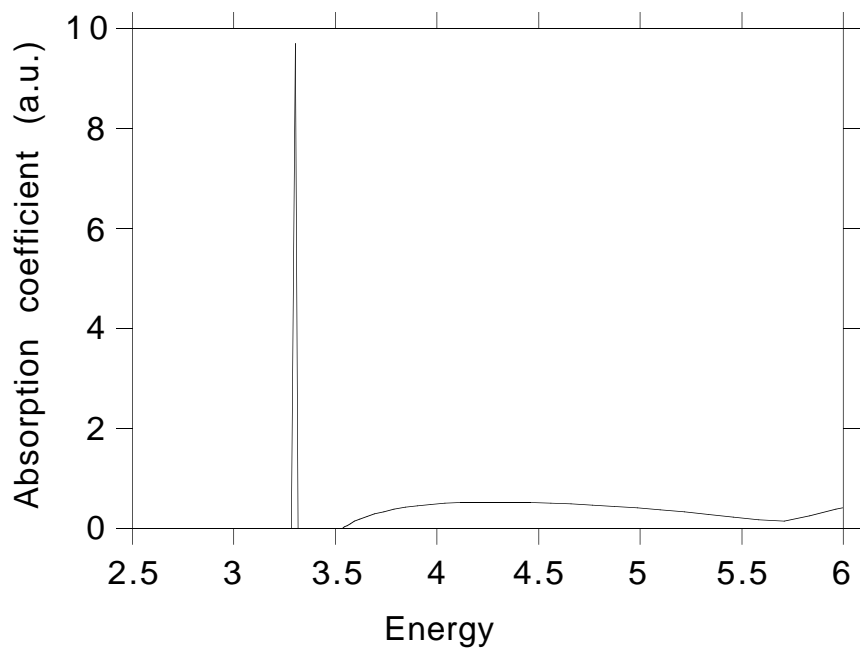

Figure 6